\documentclass[english,aps,pra,reprint,noshowpacs,superscriptaddress]{revtex4-1}   
\usepackage[T1]{fontenc}	
\usepackage[latin9]{inputenc}	
\usepackage{geometry}		
\usepackage{graphicx}
\usepackage[above,below]{placeins}	
\usepackage{times}
\usepackage{hyperref}  
\hypersetup{colorlinks=true,urlcolor=blue,citecolor=blue,linkcolor=blue}   
\begin{document}

\title{Who does what now? How physics lab instruction impacts student behaviors}
\author{Katherine N. Quinn}
\author{Kathryn L. McGill}
\author{Michelle M. Kelley}
\author{Emily M. Smith}
\author{N. G. Holmes}
\affiliation{Laboratory of Atomic and Solid State Physics, Physics Department, Cornell University, 144 East Ave, Ithaca, NY, 14850}


\begin{abstract}
While laboratory instruction is a cornerstone of physics education, the impact of student behaviours in labs on retention, persistence in the field, and the formation of students' physics identity remains an open question. In this study, we performed in-lab observations of student actions over two semesters in two pedagogically different sections of the same introductory physics course. We used a cluster analysis to identify different categories of student behaviour and analyzed how they correlate with lab structure and gender. We find that, in lab structures which fostered collaborative group work and promoted decision making, there was a task division along gender lines with respect to laptop and equipment usage (and found no such divide among students in guided verification labs).
\end{abstract}

\maketitle

\section{Introduction}
\vspace{-0.1in}
There is a large gender disparity in representation in the physics community, with men dominating in both rank and number~\cite{Pettersson2011}. In studying this, much emphasis in physics education research focuses on gender gaps in \textit{performance}, such as concept inventories and course grades~\cite{Scherr,Madsen2013}. However, \textit{participation} in the physics community through the roles people take on (and in particular doing lab work) can heavily shape one's identity as a physicist~\cite{Irving2015,Irving2016}. Correspondingly, a gendered division of roles influences the modern practice of physics to be laden with masculine connotations~\cite{Gonsalves2016}. Understanding how these gendered roles develop and how they are shaped through behaviors in labs is critical. 

In this paper, we explore student participation through the \textit{behaviours} they take on in an introductory physics lab course. Previous work has shown mixed results with regards to gendered action in first-year physics labs~\cite{Danielsson2009,Jovanovic1998} such as men using desktop computers more than women~\cite{Day2016} and that management of equipment apparatus is heavily impacted by gender in mixed-group pairs~\cite{Holmes2014}. In this paper, we begin to explore the replicability and generalizability of these studies, as well as understand underlying mechanisms and implications. 

We performed a cluster analysis to categorize student behaviours, a \textit{person-centered approach} which can account for non-linearities missed in common regression analyses~\cite{Corpus2014}. We found that, in the inquiry lab sections designed to foster collaborative work and promote student agency, women used laptops and personal devices more than men, and men used lab equipment more than women. We found no such difference in traditional lab sections, in which students were guided through experiments and individually filled out worksheets. We conjecture that students in the inquiry labs were afforded the opportunity to divide tasks within their groups, and therefore did so along gendered lines. We use these results to guide future work which will aim to explore the mechanisms for this observed gender-based behaviour difference in labs.

\vspace{-0.2in}
\section{Methods}

Participants were students enrolled in the honours-level mechanics course of a calculus-based physics sequence. During Fall 2017, all students in this study attended the same lecture, were mixed together in discussion sections, but were separated into two pedagogically different lab types (three \textit{traditional lab} sections and two \textit{inquiry lab} sections). Students self-selected into their lab sections prior to the start of the course; at the time of selection, they were unaware of differences between the labs. During Spring 2018, the two lab sections under study were both inquiry labs.

The \emph{traditional labs} were designed to reinforce physics content knowledge by providing students with hands-on experiences with physical phenomena. Students were provided with a detailed lab worksheet that guided them through experiments that demonstrated physics concepts. Each student handed in their individual worksheet at the end of the lab period.

The \emph{inquiry labs} were designed to emphasize the process of experimentation in physics. In these labs, students were provided with a goal but were not provided with specific procedures or decisions for reaching that goal. Experimentation skills were emphasized in all lab activities with a focus on iterating, improving, and extending investigations. Students worked collaboratively on electronic lab notes to document their processes and submitted one set of notes per group at the end of the lab session.

\vspace{-0.15in}
\subsection{Quantifying student behaviours}

In all lab sections, observers documented student behaviours following the observation protocol used in Day \textit{et. al.}~\cite{Day2016}. Every five minutes, an observer noted each student's actions in the lab using the codes explained in Table~\ref{table:codes}. The cumulative actions of a student in a given lab period formed a student profile. Thus over the course of a semester there are multiple profiles for each student, one for each lab period. A profile is constructed by normalizing the frequency of observed codes for a student in a lab period (and therefore represents the fraction of codes associated with each student). In Fall 2017, observers were physically present in the lab space. In Spring 2018, observers coded video using the same protocol to determine student profiles.

\begin{table}[htbp]
\caption{\textbf{Action codes used in observations}. The \textit{Laptop} code is used for both handling a laptop or personal device (students used laptops, phones, and tablets for the purpose of notetaking, writeup, data analysis and reading instructions in the inquiry labs). \label{table:codes}}
\begin{ruledtabular}
\begin{tabular}{cl}
\textbf{Code} & \textbf{Description} \\
 \hline
Equipment & Handling equipment \\
Laptop & Using a laptop or personal device \\
Paper & Writing on paper or in a notebook \\
Computer & Using the desktop computer at the lab bench \\
Other & Other behaviour, such as discussing or observing
\end{tabular}
\end{ruledtabular}
\vspace{-0.1in}
\end{table}

Codes were applied by identifying what the students handled (laptop or personal device, computer, paper, equipment). The \textit{Other} code captured all other actions such as talking with peers, asking questions of the instructor, listening to explanations, observing other group members, and off-task behaviour. One code was applied to each student in the class every five minutes, except in cases where the student had not yet arrived, had already left, or could not be easily identified (such as walking off camera).

To validate this method, two observers coded student actions in the same lab period using the described protocol but at different five-minute intervals to independently determine student profiles. Observers were not coding the same student at the same time. This was done to address two issues: (1) the reliability of the codes, and (2) the validity of the five minute time interval at capturing overall student behaviours in a two-hour lab period. A chi-squared analysis was performed on the contingency table constructed from the cumulated student profiles (frequencies of each code). In all cases observers' profiles were not significantly different ($p > 0.1$). Because each pair of observers obtained statistically indistinguishable observations, single observers coded subsequent lab periods.

Through in-class surveys, students self-reported demographic information. In all, 143 students were used in this study, resulting in 522 student profiles across 30 lab periods (each student is assigned a unique profile per lab period, and is in at most 5 lab periods). Table~\ref{table:Demographics} shows the gender demographics of the two lab sections. Although surveys provided students with the opportunity to disclose another gender, no student chose to do so.

To compare profiles from all students across all lab sections in both semesters, each student profile was normalized so that each measure represented the fraction of codes rather than the number of codes (see Table~\ref{table:codes} for list of codes). To perform a cluster analysis, each profile was grand mean scaled (Mean~=~0, SD~=~1), thus turning the different measures into z-scores~\cite{Schmidt2017,Corpus2014}. The Euclidean distance between student profiles represents the dissimilarity of student profiles, in units of standard deviations~\cite{Corpus2014}. In this way, we can relate geometric quantities (Euclidean distances) to statistical quantities (dissimilarities between profiles).

\begin{table}[htbp]
\caption{\textbf{Student demographics} of this study, with numbers in paranthises. In all, 143 students were used in this study. Students were observed during multiple lab classes during the semester, resulting in 522 student behaviour profiles. \label{table:Demographics}}
\begin{ruledtabular}
\begin{tabular}{l c c c c}
 & \multicolumn{2}{c}{\textbf{Traditional Labs}} & \multicolumn{2}{c}{\textbf{Inquiry Labs}} \\
					& Students & Profiles & Students & Profiles \\
					& $\%(N)$ & $\%(N)$ & $\%(N)$ & $\%(N)$\\
 \hline
Women & $19 \pm 5 (11)$ & $18\pm 3(34)$ & $25\pm 5 (21)$ & $26\pm 2(87)$ \\
Men & $79\pm 5(46)$ & $81 \pm 3(152)$ & $74 \pm 5 (63)$ & $74 \pm 2(226)$ \\
Undisclosed & $2\pm2 (1)$ & $1\pm 1(2)$ & $1\pm 1 (1)$ & $0.3\pm 0.3(1)$
\end{tabular}
\end{ruledtabular}
\vspace{-0.1in}
\end{table}

\vspace{-0.2in}
\subsection{Cluster analysis}

\begin{figure}
\includegraphics[width=0.9\linewidth]{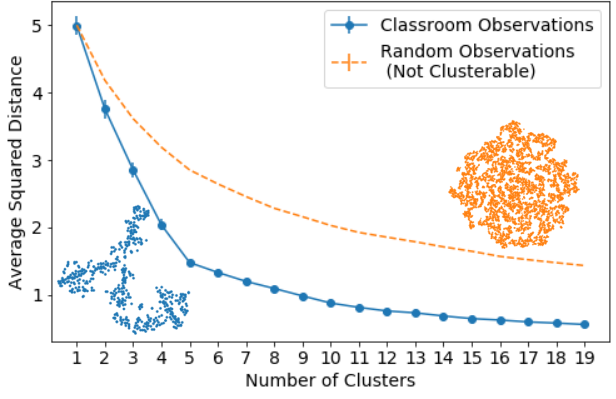}
\caption{Average squared distance from each point to the center of its assigned cluster, illustrating the use of the elbow method~\cite{elbow} to determine the optimal number of clusters. Blue points represent clustered student profiles, and orange points represent ten thousand randomly generated (non-clusterable) points for comparison. Points are illustrated using a t-SNE visualization~\cite{VanDerMaaten2008} for qualitative comparison, with random points forming a blob and classroom observations showing structure.\label{fig:elbow}}
\end{figure}

\vspace{-0.1in}Once all student profiles were obtained in z-score format, a standard k-means cluster analysis was performed. K-means is an iterative algorithm, where the optimal solution is found when the sum of square of distances from all points to their respective cluster center is minimized~\citep{kMeans}. We used the elbow method to determine if the data are clusterable~\cite{elbow}. This method optimizes the number of clusters by looking at the square distance from each point to their respective cluster center, and plotting this as a function of the number of clusters. When averaged over the number of profiles, it represents the variance of the data. Note that increasing the number of clusters will always decrease the average squared distance, because allowing more clusters will explain more variance in the data. The ``elbow" in the plot corresponds to the optimal number of clusters. Fig.~\ref{fig:elbow} illustrates the method and compares to random (unclusterable) data. From this we determined the optimal number of clusters to be five (a coincidence of this study, and not a reflection of the number of unique codes, as the random data does not have an elbow at five). The clusters account for 70\% of the variance in the data (64\% of equipment use, 78\% of paper and notebook use, 79\% of laptop and personal device use, 73\% of lab desktop computer use, and 59\% of other activities), well above the 50\% threshold used for a study of this type~\cite{Schmidt2017,Corpus2014}.

\begin{figure}
\vspace{-0.1in}
\includegraphics[width=\linewidth]{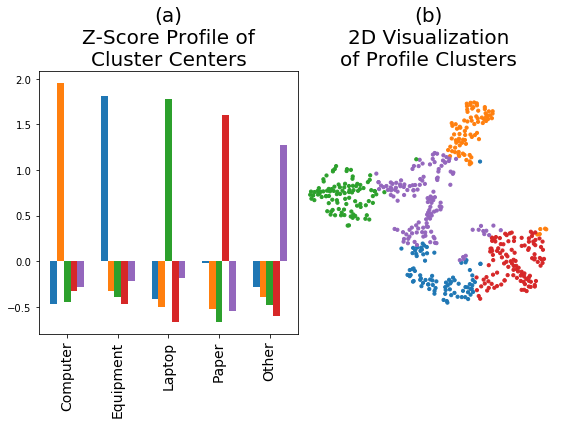}
\caption{(a) The z-score profile of each cluster center shows a division based on task, and so we name the clusters according to the codes from Table~\ref{table:codes}. (b) Student profiles are visualized in two dimensions using t-SNE and colored by cluster. Since this image is an attempt to project a five dimentional space into two dimensions, it provides a \textit{qualitative} picture of the cluster shapes~\cite{VanDerMaaten2008}.\label{fig:ZScores}}
\vspace{-0.1in}
\end{figure}

Clusters are primarily characterized by their centers, and so we label each cluster based on a description of their respective center. In Fig.~\ref{fig:ZScores}{(a)}, we see that the centers correspond to the five codes described in Table~\ref{table:codes}, i.e. a student profile in the equipment cluster corresponds to a strong positive deviation from the average equipment use. In Fig.~\ref{fig:ZScores}{(b)} we use t-stochastic network embedding (t-SNE)~\cite{VanDerMaaten2008} to visualize the clusters, where each point in the figure represents an individual student profile. Since we are attempting to project a five-dimensional space into two-dimensions, the resulting image primarily preserves structure and is used for a qualitative visualization, with distant points dissimilar and close points similar to each other.

\vspace{-0.2in}
\section{Results}

\begin{figure}
\vspace{-0.1in}
\includegraphics[width=\linewidth]{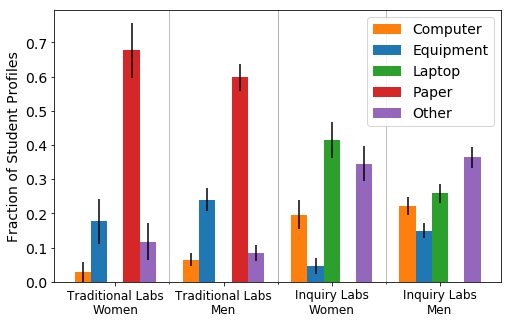}
\caption{Composition of each cluster. The biggest difference occurs in laptop and paper usage, which reflects the different logistical differences in the traditional and inquiry labs. While there is no statistically significant difference in the distribution of men's and women's profiles in the traditional labs ($p=0.65$), there \textit{is} a difference in inquiry labs ($p=0.011$), specifically with regards to equipment and laptop usage. Label colours match cluster colours from Fig.~\ref{fig:ZScores}.\label{fig:clusterComposition}}
\end{figure}

Once the student profiles were clustered, we analyzed each cluster's composition, shown in Fig.~\ref{fig:clusterComposition}. Profiles from students who chose not to disclose their genders (n=2) are omitted from this part of the analysis. The first striking difference in cluster composition is that the \textit{Laptop} cluster is composed entirely of students in the inquiry labs, and that the \textit{Paper} cluster is composed entirely of students in the traditional labs. This reflects the logistical differences between these two sections. In the traditional labs, students wrote answers to prompts on paper worksheets. In contrast, students in the inquiry labs worked collaboratively on electronic lab notes, and so documented everything using electronic devices (laptops, personal devices, or the desktops provided in lab). In both the traditional and inquiry labs, students needed to use equipment and were provided with a desktop computer. Therefore, as expected, the \textit{Computer} and \textit{Equipment} clusters contain students from both the traditional and inquiry labs. We note here, and will discuss further in the Section~\ref{sef:disc}, that these codes reflect \textit{what} a student was handling and not \textit{why} they were handling it (i.e. a desktop computer can be used for data gathering, analysis or writeup).

Because each student has multiple profiles arising from the different lab sessions throughout the semester, we analyzed whether or not individual students' profiles appear in multiple clusters over the semester. In the traditional labs, $87\pm 4\%$ of students have profiles in more than one cluster. Similarly, in the inquiry labs $86\pm 4\%$ of students have profiles in more than one cluster. This suggests that student profiles cannot be further collapsed to indicate `semester long' behaviour, since they vary from week to week (for various reasons, such as variability in lab content and students changing lab partners).

Figure~\ref{fig:clusterComposition} shows that, in the traditional labs, there is no statistically significant difference in the fraction of men's and women's profiles in any of the clusters ($p=0.65$). However, we notice that there \textit{is} a difference in the cluster composition with respect to men's and women's profiles in the inquiry labs ($p=0.011$). Specifically, $44\pm6\%$ of women's profiles in the inquiry labs are in the \textit{Laptop} cluster compared to $25\pm 3\%$ of men's profiles, and $4\pm 2\%$ of women's profiles are in the \textit{Equipment} cluster compared to $14\pm2\%$ for men. This suggests a division of tasks along gender lines in the inquiry labs, with women using laptops and personal devices more than men and men using equipment more than women.

\begin{table}[htbp]
\caption{\textbf{Student averaged fraction of codes} in the inquiry labs, for handling equipment or using a laptop or personal deivce, broken down by gender. All other code comparisons have $p>0.2$, indicading no statistically significant difference.\label{table:timeFrac}}
\begin{ruledtabular}
\begin{tabular}{l c c c}
 & \textbf{Women} & \textbf{Men} & \textbf{p Value}\\
 \hline
\textbf{Equipment} & $9\pm 1\%$ &  $13\pm1\%$ & 0.0097\\
\textbf{Laptop} & $31\pm3\%$ & $23\pm 2\%$ & 0.0082 
\end{tabular}
\end{ruledtabular}
\end{table}

We looked at average fraction of codes in the inquiry lab to see if the results corroborate or refute the results of the cluster analysis. Table~\ref{table:timeFrac} shows these averages, broken down by gender. These results support our results from the cluster analysis. Men spent a larger fraction of their coded time handling equipment than women, and women spent a larger fraction of their coded time on a laptop or personal device than men.

\vspace{-0.1in}
\section{Discussion and Conclusions}
\label{sef:disc}

In this study, we analyzed the in-lab behaviours of students in two pedagogically different lab sections of the same introductory physics course. The biggest effect impacting the cluster composition was due to logistical differences between the traditional and inquiry labs (with regards to laptop and paper usage), an expected result given the large structural differences between labs. Furthermore, we found a second-order effect with respect to gender. We found no gendered difference in the traditional labs, but we did find one in the inquiry labs. We conjecture that this is because students in the inquiry labs were afforded the opportunity to divide tasks within their groups and did so along gendered lines.

Students in the traditional labs worked in groups to closely follow detailed lab worksheets but individually filled in answers. In other words, they completed a specific (assigned) individual task within a group setting where they shared equipment. However, there was very little room for decision making, as they followed specific instructions designed to demonstrate physics concepts. In contrast, students in the inquiry labs needed to decide as a group how to meet the provided goal of the lab and submitted one electronic lab notebook per group rather than individual lab worksheets. The inquiry labs were designed to foster collaboration within and between groups so students were free to divide tasks, and did so along gendered lines (with men handling equipment more than women, and women handling laptops and personal devices more than men).

These results raise many questions about equity in lab groups, in particular (1) how gendered roles are constructed and (2) how tasks are assigned. Studying more nuanced yet conceptually different tasks (such as data analysis versus secretarial note-taking) could provide insight into the mechanism behind gendering roles in lab classes. There is high variability with regards to the specifics of lab sections, which can lead to different amounts of equipment and laptop usage. For example, our results seem to contradict previous work which showed men using desktop computers more than women~\cite{Day2016}. Instead, analyzing \textit{why} a student is engaging in a particular task (such as using a computer for secretarial notetaking versus for data analysis) could provide a deeper understanding of student behaviour. Video recordings of individual groups were captured during the course of this study, and will be analyzed to answer specific questions with regards to task allocation. In future work, we will also evaluate the impacts of adding structure to the group work in the inquiry labs, such as deliberately assigning students to roles based on concepts such as those used in cooperative grouping~\cite{Heller1992}. In this way, students can maintain agency in decision making with respect to experimental design and data analysis, while structure is provided for role assignments. We plan to compare the behaviour of students in such labs to the ones used in this study and see if a gender-based behaviour difference persists.

\vspace{-0.1in}
\acknowledgments{We thank the teaching assistants and lab instructors for the course used in this study for their invaluable support and cooperation.This study was supported by the President's Council for Cornell Women's Affinito-Stewart Grant and the Cornell's College of Arts and Sciences Active Learning Initiative.}

\bibliographystyle{apsrev}  	
\bibliography{perc2018_refs}  	

\end{document}